\documentclass[aps,pre,twocolumn,showpacs,superscriptaddress]{revtex4}
\usepackage[T2A]{fontenc}
\usepackage[cp1251]{inputenc}
\usepackage[english]{babel}
\usepackage{graphicx}
\usepackage{color}
\usepackage{bm}
\usepackage{amssymb}
\usepackage[intlimits]{amsmath}
\usepackage{amsmath}

\newcommand{\be}{\begin{equation}}
\newcommand{\ee}{\end{equation}}
\newcommand{\ben}{\begin{equation*}}
\newcommand{\een}{\end{equation*}}

\begin{document}

\title{Next-to-leading order corrections to  capacity for nondispersive nonlinear optical fiber channel in intermediate power region}

\author{A.~A.~Panarin}
\email[Electronic address: ]{panarin.a.a@mail.ru} \affiliation{Novosibirsk State
University, Novosibirsk, 630090 Russia}
\author{A.~V.~Reznichenko}
\email[Electronic address: ]{a.v.reznichenko@inp.nsk.su} \affiliation{Budker
Institute of Nuclear Physics of Siberian Branch Russian Academy of Sciences,
Novosibirsk, 630090 Russia} \affiliation{Novosibirsk State University, Novosibirsk,
630090 Russia}
\author{I.~S.~Terekhov}
\email[Electronic address:]{i.s.terekhov@gmail.com} \affiliation{Budker Institute of
Nuclear Physics of Siberian Branch Russian  Academy of Sciences, Novosibirsk, 630090
Russia}
 \affiliation{Novosibirsk State University, Novosibirsk, 630090 Russia}

%\author{
%\IEEEauthorblockN{A.~A.~Panarin\IEEEauthorrefmark{1}\IEEEauthorrefmark{2},
%A.~V.~Reznichenko\IEEEauthorrefmark{1}\IEEEauthorrefmark{2},
%I.~S.~Terekhov\IEEEauthorrefmark{3}}\\
%    \IEEEauthorblockA{\IEEEauthorrefmark{1}Novosibirsk State University, Novosibirsk, 630090, Russia}\\
%    \IEEEauthorblockA{\IEEEauthorrefmark{2}Budker Institute of  Nuclear
%Physics, Russian  Academy of Sciences, Novosibirsk, 630090, Russia}
%}

%=========================================Abstract

%\hskip 2cm
%\date{\today}

\begin{abstract}

We consider the optical fiber channel modelled by the nonlinear Shr\"{o}dinger
equation with zero dispersion and  additive Gaussian noise. Using Feynman
path-integral approach for the model we find corrections to  conditional probability
density function, output signal distribution, conditional and output signal
entropies, and the channel capacity at large signal-to-noise ratio. We demonstrate
that the correction to the channel capacity is positive for large signal power.
Therefore, this correction increases the earlier calculated capacity for a
nondispersive nonlinear optical fiber channel in the intermediate power region.

\end{abstract}
 \maketitle

%==================Introduction
\section{Introduction.}

The problem of information transmission through a noisy communication channel is
considered more than 60 years. First results of the solution of the problem were
obtained by Shannon Ref. \cite{Shannon:1948}. In Ref. \cite{Shannon:1948} Shannon
introduced the channel capacity $C$,  which gives the maximum amount of information
that can be reliably transmitted over a noisy communication channel. For the first
time he obtained the logarithmic dependence of the capacity on signal power for a
linear communication channel with additive Gaussian noise:
\begin{eqnarray}\label{CapacityShannon}
C\propto\log\left(1+\mathrm{SNR}\right)\,,
\end{eqnarray}
where $\mathrm{SNR}=P/N$ is the signal-to-noise power ratio, $P$ is the signal
power, and $N$ is the noise power. It means that in order to increase the capacity
one has to increase the signal power $P$ for the fixed noise power $N$. There is a
question: how the nonlinearity in a communication channel affects the result
(\ref{CapacityShannon}). The interest to the nonlinear channels started to increase
when the fiber optics communication system began intensively developing. It is
connected with the Kerr nonlinearity in optical fibers. The influence of
nonlinearity on capacity is investigated both for dispersive and nondispersive
optical channels. The channels with dispersion were studied in numerous papers, see,
e.g.,
\cite{Mitra:2001,Narimanov:2002,Kahn:2004,Essiambre:2008,Essiambre:2010,Killey:2011,Agrell:2014,
Sorokina2014,Terekhov:2016b,Terekhov:2016c,Turitsyn:2012} and references therein.
Despite the fact that the capacity for the nonlinear channel with dispersion was
considered  in many papers the exact in nonlinearity result is still not found due
to difficulty of the problem. Therefore as the first step in understanding of the
effects of nonlinearity impact in the channel one can consider the nonlinear channel
with zero average dispersion.  The nonlinear nondispersive optical fiber channels
are also considered in numerous papers, see, e.g.,
\cite{M1994,MS:2001,Tang:2001,tdyt03,Mansoor:2011,Terekhov:2016a}. Of course, the
problem of capacity calculation for these channels is simpler than the problem with
dispersion. However it is still quite a challenging  problem especially at large
parameter $\mathrm{SNR}$, and new techniques and methods are highly desirable to
advance these studies
\cite{Narimanov:2002,tdyt03,Mansoor:2011,Agrell:2012,Agrell:2013,Terekhov:2014}.

%===========================NOTIONS
The channel capacity $C$ can be determined as the maximum of the mutual information
$I_{P_{X}[X]}$ with respect to the probability density function (PDF) $P_{X}[X]$ of
an input signal $X$:
\begin{eqnarray}
C=\max_{P_{X}[X]} I_{P_{X}[X]} .\label{capacity1}
\end{eqnarray}
The  maximum value of the mutual information $I_{P_{X}[X]}$ in Eq.~(\ref{capacity1})
should be found at the given average signal power:
\begin{eqnarray} \label{power}
P=\int {\cal D} X  P_{X}[X] |X|^2.
\end{eqnarray}
The PDF $P_{X}[X]$ also obeys the normalization condition:
\begin{eqnarray}\label{NormPx}
\int {\cal D} X  P_{X}[X]=1,
\end{eqnarray}
that fixes the integration measure ${\cal D} X=d \mathrm{Re} X d \mathrm{Im} X$. The
mutual information is defined as the difference of  output signal  entropy $H[Y]$
and conditional entropy $H[Y|X]$:
\begin{eqnarray}
& I_{P_{X}[X]}= H[Y]-H[Y|X],
\label{MI}
\end{eqnarray}
where the entropies are defined  as
\begin{eqnarray}
 \!\!\! H[Y|X]&=&-\int {\cal D}X {\cal D}Y P_{X}[X]{P[Y| X]} \log P[Y| X],
\label{condentropy}\\
\!\!\! H[Y]&=&-\int{\cal D}Y P_{\mathrm{out}}[Y] \log P_{\mathrm{out}}[Y],
\label{entropies}
\end{eqnarray}
here $P_{\mathrm{out}}[Y]$ is output signal PDF:
\begin{eqnarray}
\!\!\! P_{\mathrm{out}}[Y]=\int {\cal D}X P_{X}[X] P[Y|X], \label{Pout}
\end{eqnarray}
and $P[Y|X]$ is conditional probability density function, i.e., the PDF to have the
output signal $Y$ when the input signal is $X$. The measure ${\cal D}Y$ is defined
as $\int {\cal D}Y P[Y|X] =1$. Our  definitions (\ref{MI})--(\ref{entropies}) imply
that we measure the capacity in units $(\log 2)^{-1}$ bit per symbol (also known as
nat per symbol). Usually the input and output signals are the functions of time
which have certain bandwidth. Therefore the sampling of the temporal signal should
be introduced to define discrete-time memoryless channel. In this case the capacity
should be proportional to bandwidth. But we discuss nondispersive channels. It means
that we can consider the functions $X(t)$ and $Y(t)$ at the same time moment and
calculate only per-sample (i.e., for one time elementary channel) quantities.

To calculate the mutual information we should know the conditional probability
density function $P[Y|X]$ for the channel. This quantity depends on the channel
model. As was mentioned above for the nondispersive channel the temporal signal
waveform changes during propagation independently for every time moment. Therefore,
instead of consideration of the evolution of $\psi(z,t)$ we can consider a set of
parallel independent scalar channels \cite{M1994, Mansoor:2011}, the so-called
per-sample channels. We choose the signal propagation model described by the
following equation, see \cite{Mansoor:2011}:
\begin{align}
& \partial_{z}\psi(z)-i\gamma |\psi(z)|^2 \psi(z)=\eta(z),
\label{Shrodingerequation}
\end{align}
i.e., the nonlinear Shr\"{o}dinger equation with zero  dispersion and with additive
noise $\eta(z)$. In Eq.~(\ref{Shrodingerequation}) $\psi(z)$ is the complex function
which describes the signal propagation in the channel, $\gamma$ is Kerr nonlinearity
parameter, the function $\eta(z)$ describes the additive noise in the channel. The
noise  has the zero mean $\langle \eta (z) \rangle_{\eta} = 0$ and the correlation
function $\langle \eta (z)\bar{\eta}(z') \rangle_{\eta} = Q \delta(z-z^\prime)\,$,
where $Q$ is the noise power per unit length. The function $\psi(z)$ obeys the
boundary condition $\psi(0)=X$. In our notations the per-sample signal power and
noise power are $P$ and $N=QL$, respectively, where $L$ is the signal propagation
length. Here the signal power $P$ is defined in Eq.~(\ref{power}). For the channel
(\ref{Shrodingerequation}) the conditional PDF $P[Y|X]$, i.e., the probability
density to receive the signal $\psi(L)=Y$ when $\psi(0)=X$, was found in the form of
infinite series \cite{tdyt03, M1994} within Martin-Siggia-Rose formalism based on
the quantum field theory methods \cite {MSR:1973, Zinn-Justin}. Using the obtained
probability $P[Y|X]$ the lower bound for the channel capacity at large
$\mathrm{SNR}=P/(QL)$ was found:
\begin{eqnarray}
\label{Capacityboundary} \!\!\!C \geq \!\frac{\log\left(\mathrm{SNR}\right)}{2}+
\frac{1+\gamma_{E} -\log (4 \pi) }{2}+{\cal O}\!\left(\!\frac{\log (\mathrm{SNR})}{
\mathrm{SNR}}\!\right)\!\!,
\end{eqnarray}
where $\gamma_{E} \approx 0.5772$ is the Euler constant. The first term in the
right-hand side of the inequality ~(\ref{Capacityboundary}) was obtained in
Ref.~\cite{tdyt03}, whereas the second term was obtained in
Ref.~\cite{Terekhov:2016a}. One can see that the lower bound
(\ref{Capacityboundary}) of the capacity grows as $(1/2)\log (\mathrm{SNR})$ instead
of $\log \mathrm{SNR}$. The factor $1/2$ appears due to the loss of information
about the phase of the signal, see Ref.\cite{Mansoor:2011}. In Ref.~
\cite{Terekhov:2016a} the new method of calculation of the conditional PDF $P[Y|X]$
was developed. This method allowed us to sum the infinite series for $P[Y|X]$
obtained in Refs. \cite{tdyt03,M1994} at large $\mathrm{SNR}$, and to obtain the
simple form of the conditional PDF $P[Y|X]$ in the leading order in
$1/\mathrm{SNR}$, see Ref. \cite{Terekhov:2016a}.  In Ref. \cite{Terekhov:2016a}
using this form of $P[Y|X]$ we calculated the capacity of the nonlinear
nondispersive optical fiber channel in the intermediate power region
\begin{eqnarray}\label{region1}
QL\ll P \ll \left(Q \gamma^2 L^3 \right)^{-1}
\end{eqnarray}
with the accuracy ${\cal O}(QL/P)+{\cal O}(\gamma^2 Q P L^3 )$. Moreover, it was
shown that at sufficiently large power $P$ in the region
\begin{eqnarray}\label{region1}
(\gamma L)^{-1}\ll P \ll \Big(Q \gamma^2 L^3 \Big)^{-1}
\end{eqnarray}
the found capacity is greater than the bound  (\ref{Capacityboundary}), but in the
region the capacity grows only as $\log\log P$ with increasing of signal power $P$
instead $(1/2)\log \mathrm{SNR}$, see Eq.~(54) in Ref.~\cite{Terekhov:2016a}.
However at $P\gg (Q \gamma^2 L^3)^{-1}$ the capacity should be of the order of
$(1/2)\log (P/QL)$. It means that we have to understand how one asymptotical regime
for the capacity transforms to another one. To this end we should calculate the
first nonzero corrections in parameter $Q L$. Moreover, to clarify the accuracy of
the results obtained in Ref.\cite{Terekhov:2016a} we also should find the first
nonzero correction to the channel capacity which is proportional to the noise power
$Q L$. To calculate the correction to the channel capacity $C$ we should know the
corrections of this order to the conditional PDF $P[Y|X]$, entropies
(\ref{condentropy})--(\ref{entropies}), and the optimal input signal distribution
$P_{\mathrm{opt}}[X]$.

The paper is organized in the following way. In Sec. II we present the results of
calculations of the next-to-leading order correction to the conditional PDF
$P[Y|X]$. In this Section we briefly remind the method of $P[Y|X]$ calculation
developed in details in Ref.~\cite{Terekhov:2016a}. The result of the calculation of
the output signal distribution $P_{\mathrm{out}}[Y]$ in the next-to-leading order
concludes Section II. Sec. III is devoted to the calculation of the conditional
entropy $H[Y|X]$ and the output signal entropy $H[Y]$ in the next-to-leading order
in $1/\mathrm{SNR}$. In Sec. IV we present the calculation of the optimal input
signal distribution $P_{\mathrm{opt}}[X]$, and in  Sec. V using the obtained
expression for $P_{\mathrm{opt}}[X]$ we find the correction to the capacity
(\ref{capacity1}). We discuss our results in Sec. VI.

%==========================================Conditional probability
\section{Calculation of the conditional PDF $P[Y|X]$ and output signal PDF $P_{\mathrm{out}}[Y]$ at large $\mathrm{SNR}$}
%=================
\subsection {Method for the conditional PDF $P[Y|X]$ calculation}
This section is based on the method described in details in Ref.
\cite{Terekhov:2016a}, therefore here we just schematically describe the
calculation. We start our consideration from the expression for the conditional PDF
$P[Y|X]$ in the path-integral form  \cite{tdyt03,Zinn-Justin, Feynman} in retarded
discretization scheme, see, e.g., Supplemental Materials  of
Ref.~\cite{Terekhov:2014} or Ref.~\cite{Terekhov:2016a}:
\begin{eqnarray} \label{ProbabInitial}
&P[Y|X] = \!\!\!\! \int\limits_{\psi(0)=X}^{\psi(L)=Y}  \!\!\!\! {\cal D}\psi
\exp\Big\{-\frac{S[\psi]}{Q}\Big\} \,,
\end{eqnarray}
where the effective  action $S[\psi]$ reads
$$S[\psi]= \int\limits_{0}^{L}dz \Big|\partial_{z}\psi-i\gamma |\psi|^2 \psi\Big|^2.$$
In the case when the parameter $\mathrm{SNR}\gg 1$ it is convenient to rewrite the
form (\ref{ProbabInitial}) in the following way, see Ref.~\cite{Terekhov:2014}:
\begin{eqnarray}
\label{QuasiclassProbabInitial} P[Y|X] =\Lambda e^{-\frac{S[\Psi_{cl}(z)]}{Q}}\,,
\end{eqnarray}
where the normalization factor is
\begin{eqnarray}\label{Normalization}
\Lambda=\int\limits_{\tilde{\psi}(0)=0}^{\tilde{\psi}(L)=0}\!\!\!\!\! {\cal
D}\tilde{\psi} \,e^{-\frac{S[\Psi_{cl}(z)+\tilde{\psi}(z)]-S[\Psi_{cl}(z)]}{Q}},
\end{eqnarray}
and the function $\Psi_{cl}(z)$ is the ``classical'' solution of the equation
$\delta S[\Psi_{cl}]=0$, where $\delta S$ is the variation of the action $S[\psi]$.
The equation for the function $\Psi_{cl}$ can be written in the form:
\begin{eqnarray}
\label{classicalTrajectoryEq} \frac{d^2\Psi_{cl}}{dz^2}-4i\gamma
\left|\Psi_{cl}\right|^2 \frac{d\Psi_{cl}}{dz}-3\gamma^2 \left|\Psi_{cl}\right|^4
\Psi_{cl}=0,
\end{eqnarray}
with the boundary conditions $\Psi_{cl}(0)=X$, $\Psi_{cl}(L)=Y$. To calculate the conditional probability we should calculate the exponent contribution and the path-integral in Eq. (\ref{QuasiclassProbabInitial}).

We start our calculation from exponent $e^{-\frac{S[\Psi_{cl}(z)]}{Q}}$. Since we
calculate the function $P[Y|X]$ with the accuracy $1/\mathrm{SNR}$ we  should find
the solution of Eq. (\ref{classicalTrajectoryEq}) with this accuracy. Following Ref.
\cite{Terekhov:2016a} we find such solution linearizing Eq.
(\ref{classicalTrajectoryEq}) in the vicinity of the solution $\Psi_0(z)$ of the
channel equation (\ref{Shrodingerequation}) with zero noise. The function
$\Psi_0(z)$ reads
\begin{eqnarray}
\label{zeronoisesolution} \Psi_0(z)=\rho \exp\left\{i \mu \frac{z}{L}+ i
\phi^{(X)}\right\},
\end{eqnarray}
where $\mu= \gamma L |X|^2$. Note that the solution (\ref{zeronoisesolution}) is
also solution of  Eq. (\ref{classicalTrajectoryEq}) but it satisfies only the input
boundary condition $\Psi_0(0)=X=\rho \,e^{i\phi^{(X)}}$, where $\rho=|X|$.
Therefore, to fulfill the output boundary condition $\Psi_{cl}(L)=Y$ we look for the
solution of Eq. (\ref{classicalTrajectoryEq}) in the form
\begin{eqnarray} \label{psiclasskappa}
\Psi_{cl}(z)=\Big(\rho+\varkappa(z)\Big)\exp\left\{i \mu \frac{z}{L}+i
\phi^{(X)}\right\},
\end{eqnarray}
where  the function $\varkappa(z)$ is assumed to be small: $|\varkappa(z)| \ll
\rho$. In Ref.~ \cite{Terekhov:2016a} we argued that statistically significant for
$P[Y|X]$ functions $\varkappa(z)$ are at least of the order of $\sqrt{Q }$. The
equation for the function $\varkappa$ has the form, see Eq.~(79) in
Ref.~\cite{Terekhov:2016a}:
\begin{widetext}
\begin{eqnarray}
\label{kappaequationexact}
&&\frac{d^2\varkappa}{dz^2}-2i\frac{\mu}{L}\frac{d\varkappa}{dz}-4 \frac{\mu^2}{L^2}
\mathrm{Re}[\varkappa]= 4i \frac{\mu}{L \rho}\left(\varkappa+\bar{\varkappa}\right)
\frac{d\,\varkappa}{dz}+ \frac{ \mu^2}{L^2
\rho}\left[5\varkappa^2+10|\varkappa|^2+3\bar{\varkappa}^2\right]+\nonumber
\\&& \frac{|\varkappa|^2 \mu}{L^2 \rho^2}\left[ 4 i L
\frac{d\varkappa}{dz}+9{\mu}\bar{\varkappa}+ 14{\mu}{\varkappa}\right]+\frac{3
\mu^2}{L^2 \rho^2}\varkappa^3+ \frac{3 \mu^2}{L^2
\rho^3}|\varkappa|^2\left[3|\varkappa|^2+2\varkappa^2\right]+ \frac{3 \mu^2}{L^2
\rho^4}|\varkappa|^4\varkappa.
\end{eqnarray}
\end{widetext}
The boundary conditions for $\varkappa$ are as follows:
\begin{eqnarray}\label{kappaboundary}
\varkappa(0)=0,\,\varkappa(L)=Y e^{-i \phi^{(X)}-i\mu}-\rho \equiv x_0+iy_0,
\end{eqnarray}
Since the $|\varkappa|\ll \rho$ we can solve Eq. (\ref{kappaequationexact}) using perturbation theory in the parameter $\varkappa/\rho$ and present the solution $\varkappa$ in the form
\begin{eqnarray}
\varkappa(z)=\varkappa_1(z)+\varkappa_2(z)+\varkappa_3(z)+\ldots
\end{eqnarray}
The functions $\varkappa_1(z) \propto \sqrt{Q }$ and $\varkappa_2(z) \propto {Q }$
were found in Ref. \cite{Terekhov:2016a}: see Eqs. (82), (86), and (87) therein. The
equation for the function $\varkappa_3(z)$ can be easily obtained from Eq.
(\ref{kappaequationexact}). The equation for the function $\varkappa_3(z)$ and the
solution of this equation are cumbersome, therefore, we do not present them here.
But we present the final result $S[\Psi_{cl}]$ in the leading $S_1$, next-to-leading
$S_2$, and next-to-next-to-leading order  $S_3$ in parameter
$1/\sqrt{\mathrm{SNR}}$:
\begin{eqnarray}\label{ActionExpansion}
S[\Psi_{cl}]= S_1+S_2+S_3+{\cal
O}\left(\mathrm{SNR}^{-5/2}\right),
\end{eqnarray}
where
\begin{eqnarray}
S_1&=&\frac{(1+4\mu^2/3)x^2_0-2\mu x_0
y_0+y^2_0}{L(1+\mu^2/3)}\,,
\end{eqnarray}
\begin{widetext}
\begin{eqnarray}
S_2&=&\frac{\mu/ \rho}{135  L \left(1+\mu ^2/3\right)^3
}\Big\{ \mu \left(4 \mu ^4+15 \mu ^2+225\right) x^3_0+ \left(23 \mu ^4+255 \mu
^2-90\right) x^2_0 y_0+\mu  \left(20 \mu ^4+117 \mu ^2-45\right) x_0 y^2_0-\nonumber\\
&&  { 3 \left(5
\mu ^4+33 \mu ^2+30\right)   y^3_0} \Big\}\,,\\
S_3&=&
\frac{\mu ^2}{2100 L \left(\mu ^2+3\right)^5 \rho ^2}\big[x_0^4\left(148 \mu ^8-12345 \mu ^6-24570 \mu ^4-806085 \mu ^2+396900\right) -12 \mu  x_0^3 y_0\left(901 \mu
   ^6+9990 \mu ^4\right.+\nonumber\\&&
\left. 84105 \mu ^2-139860\right) +36 \mu x_0 y_0^3 \left(385 \mu ^6+6198 \mu ^4+30165 \mu^2+8820\right) -6 x_0^2 y_0^2\left(980 \mu ^8+11857 \mu ^6+24210 \mu ^4\right. -\nonumber\\
&& \left. 350595 \mu ^2-49140\right)
+3 y_0^4\left(700 \mu ^8+8365 \mu ^6+23826 \mu ^4-32535 \mu ^2-34020\right) \big]\,.
\end{eqnarray}
\end{widetext}
Since $x_0$ and $y_0$ are of the order of $\sqrt{Q}$ (see the text after Eq.~(17) in
Ref. \cite{Terekhov:2016a}) one can see that $S_1/Q$, $S_2/Q$, and $S_3/Q$ are of
the order of $(\mathrm{SNR})^0$, ${\mathrm{SNR}}^{-1/2}$, and $(\mathrm{SNR})^{-1}$,
respectively. To calculate the exponent in Eq. (\ref{QuasiclassProbabInitial}) with
the accuracy $1/\mathrm{SNR}$ we substitute the expansion  (\ref{ActionExpansion})
into the exponent and arrive at the result:
\begin{eqnarray}\label{expS}
e^{-S[\Psi_{cl}]/Q}&=& e^{-\frac{(1+4\mu^2/3)x^2_0-2\mu x_0 y_0+y^2_0}{Q L
(1+\mu^2/3)}}\left(1-\frac{S_2}{Q}+\right.\nonumber\\
&&\!\!\!\!\!\!\!\!\!\left.\left[\frac{S_2^2}{2Q^2}-\frac{S_3}{Q}\right]+{\cal O}
\left(\mathrm{SNR}^{-3/2}\right)\right).
\end{eqnarray}

To calculate the normalization factor $\Lambda$ we also use the method developed in
\cite{Terekhov:2016a}. First, we change the integration variables in
Eq.~(\ref{Normalization}) from $\tilde{\psi}(z)$ to $u(z)$ as
$\tilde{\psi}(z)=e^{i\gamma\rho^2 z}u(z)$. Then we expand
$e^{S[\Psi_{cl}(z)+\tilde{\psi}(z)]-S[\Psi_{cl}(z)]}$ in parameter $Q$, and find
terms of the order of $Q^{0}$, $Q^{1/2}$ and $Q^{1}$. After that using the Wick's
theorem and correlation function (see Eqs. (98), (103)-(105) in
Ref.\cite{Terekhov:2016a}) we obtain
\begin{eqnarray}\label{LambdaExpansion}
\!\!\!\Lambda=\frac{1}{\pi QL\sqrt{1+\mu^2/3}}\left(1+
\tilde{\Lambda}_1+\tilde{\Lambda}_2+{\cal
O}\!\left(\frac{1}{\mathrm{SNR}^{3/2}}\right)\!\!\right)\!,
\end{eqnarray}
where
\begin{widetext}
\begin{eqnarray}
\tilde{\Lambda}_1&=&- \frac{3\mu}{5\rho
(3+\mu^2)^2}\left(\mu (15+\mu^2)x_0-2(5-\mu^2/3)y_0\right), \\
\tilde{\Lambda}_2&=&\frac{\mu ^2 \left(11 \mu ^4+201 \mu ^2-504\right) Q L}{140 \left(\mu ^2+3\right)^3 \rho ^2}+\frac{\mu^2}{70 \left(3+\mu ^2\right)^4 \rho ^2} \left(\left(32 \mu ^6+453 \mu ^4+8064 \mu ^2-6237\right) x_0^2+\right.
\nonumber\\
&& \left. 12 \mu  \left(4 \mu ^4+75\mu ^2-1323\right) x_0 y_0-3 \left(7 \mu ^6+141 \mu ^4+1179 \mu ^2-567\right) y_0^2\right).
\end{eqnarray}
\end{widetext}
The correction $\tilde{\Lambda}_1$  was found in Ref.~\cite{Terekhov:2016a}, see
Eq.~(109) therein. This correction contains  $x_0$ and $y_0$ in the first power,
therefore, it is of the order of  $\sqrt{Q/\rho^2}$. The correction
$\tilde{\Lambda}_2$ contains two different terms. One term is proportional to
$Q/\rho^2$ and another one is the second order homogeneous polynomial in   $x_0$ and
$y_0$.

Using Eqs. (\ref{expS}) and (\ref{LambdaExpansion})  we obtain the expansion of the
conditional PDF:
\begin{eqnarray}\label{CondProbWithCorr}
P[Y|X]&\approx&P_0[Y|X]+\delta P_1[Y|X]+\delta P_2[Y|X]\,,
\end{eqnarray}
where
\begin{eqnarray}
\!\!\!\!\!\!\!\!P_0[Y|X]&=&\frac{e^{-\frac{(1+4\mu^2/3)x^2_0-2\mu x_0 y_0+y^2_0}{Q L
(1+\mu^2/3)}}}{\pi Q L\sqrt{1+\mu^2/3}},
\end{eqnarray}
\begin{eqnarray}
\!\!\!\!\!\!\!\!\delta
P_1[Y|X]&=&P_0[Y|X]\left(\tilde{\Lambda}_1-\frac{S_2}{Q}\right),
\end{eqnarray}
\begin{eqnarray}
\!\!\!\!\!\!\!\!\delta
P_2[Y|X]&=&P_0[Y|X]\left(\frac{S_2^2}{2Q^2}-\frac{S_3+S_2\tilde{\Lambda}_1}{Q}+\tilde{\Lambda}_2\right).
\end{eqnarray}
One can check that the conditional probability (\ref{CondProbWithCorr}) obeys the
following important properties:
\begin{eqnarray}
\lim_{Q\to 0}P[Y|X]&=&\delta\left(Y-\Psi_0(L)\right)\,,\label{limitQTo0}\\
\lim_{\gamma\to 0}P[Y|X]&=&\frac{e^{|Y-X|^2/(QL)}}{\pi QL}\,,\label{limitGammaTo0}\\
\int {\cal} D Y P[Y|X]&=&1\,.\label{IntegrationCondition1}
\end{eqnarray}
The condition (\ref{limitQTo0}) is the  deterministic limit of $P[Y|X]$ in the
absence of noise. The condition (\ref{limitGammaTo0}) means that our conditional
probability transforms to the conditional probability of the linear channel. Note
that all found corrections are proportional to the parameter $\mu=\gamma L \rho^2$,
therefore, they disappear when the nonlinearity goes to zero. The last
(normalization) condition (\ref{IntegrationCondition1})  is the check of correctness
of our calculations:  one can check that
\begin{eqnarray}\label{normalization3}
\int {\cal} D Y \delta P_{1,2}[Y|X]=0,
\end{eqnarray}
since $\int {\cal} D Y P_0[Y|X]=1$.

%=========================================
\subsection{PDF $P_{\mathrm{out}}[Y]$ of the output signal}
Now we proceed to calculation of the distribution $P_{\mathrm{out}}[Y]$ of the
output signal $Y$. Let us consider the integral, see Eq. (\ref{Pout}),
\begin{eqnarray}\label{si}
P_{\mathrm{out}}[Y]=\int {\cal D} X P[Y|X] P_{X}[X],
\end{eqnarray}
where the input signal PDF $P_{X}[X]$ is a smooth function. We assume that the
function $P_{X}[X]$ changes sufficiently when the variation of the variable $X$ is
of the order of $\sqrt{P}$. Since $QL \ll P\ll (Q L^3 \gamma^2)^{-1}$ we can
calculate the integral (\ref{si}) by the Laplace's method \cite{Lavrentiev:1987} in
the same manner as we performed the leading order calculation of
$P_{\mathrm{out}}[Y]$, see Appendix C in Ref.~\cite{Terekhov:2016a}. It is
convenient to change the integration variables from  $X=x_1+i y_1$ to $\tau= \tau_1+
i \tau_2$. The substitution has the form:
\begin{eqnarray}\label{VariableChange}
X&=&\frac{\left(\sqrt{|Y|^2-\tau_2^2}-\tau_1\right)\left(\sqrt{|Y|^2-\tau_2^2}-i \tau_2\right)}{|Y|^2}\times\nonumber\\
&&Y \exp\left\{-i\gamma L\left(\sqrt{|Y|^2-\tau_2^2}-\tau_1^2\right)^2\right\}.
\end{eqnarray}
The choice of the substitution (\ref{VariableChange}) is motivated by the  fact that
at $\tau=0$ one has $X=Ye^{-i \gamma L |Y|^2}$, and the function $P[Y|X]$ reaches
the maximum at the point $\tau=0$. After the change of variables
(\ref{VariableChange}) we perform integration using Laplace's method and obtain:
\begin{eqnarray}\label{PoutWithCorrections}
P_{\mathrm{out}}[Y]=P_X[\tilde{Y}]+\delta P_{\mathrm{out}}[\tilde{Y}],
\end{eqnarray}
where $\tilde{Y}=Ye^{-i \tilde{\mu}}=\tilde{y}_1+i \tilde{y}_2$,
$\tilde{y}_1=\mathrm{Re} \tilde{Y} $, $\tilde{y}_2=\mathrm{Im} \tilde{Y} $,
$\tilde{\mu}=\gamma L |Y|^2$. The correction $\delta P_{\mathrm{out}}[\tilde{Y}]$
can be expressed through the input signal distribution as follows:
\begin{widetext}
\begin{eqnarray} \label{deltaPout}
\delta P_{\mathrm{out}}[\tilde{Y}]&=&\frac{\gamma Q L^2}{3}\left((3\tilde{y}_2
-\tilde{\mu}\tilde{y}_1)\frac{\partial P_X[\tilde{Y}]}{\partial \tilde{y}_1}-
(3\tilde{y}_1 +\tilde{\mu}\tilde{y}_2)\frac{\partial P_X[\tilde{Y}]}{\partial
\tilde{y}_2}-\frac{1}{2}(3(\tilde{y}_1^2-\tilde{y}_2^2)+4\tilde{\mu}\tilde{y}_1\tilde{y}_2)
\frac{\partial^2 P_X[\tilde{Y}]}{\partial \tilde{y}_1\partial \tilde{y}_2}\right)+\nonumber\\
&&\frac{ Q L}{12|Y|^2}\left(\left(3 |Y|^2+6 \tilde{\mu}\tilde{y}_1\tilde{y}_2
+4\tilde{\mu}^2 \tilde{y}_2^2\right)\frac{\partial^2 P_X[\tilde{Y}]}{\partial
\tilde{y}_1^2}+\left(3 |Y|^2-6 \tilde{\mu}\tilde{y}_1\tilde{y}_2 +4\tilde{\mu}^2
\tilde{y}_1^2\right)\frac{\partial^2 P_X[\tilde{Y}]}{\partial \tilde{y}_2^2}\right).
\end{eqnarray}
\end{widetext}
In the polar coordinates $\tilde{Y}=\tilde{\rho}\,e^{i\tilde{\phi}}$ the correction
$\delta P_{\mathrm{out}}[\tilde{Y}]$ reads:
\begin{eqnarray} \label{deltaPout2}
\delta P_{\mathrm{out}}[\tilde{Y}]&=&-\frac{\gamma Q L^2 }{2}
\frac{\partial}{\partial \tilde{\phi}}\Big(1+\tilde{\rho} \frac{\partial}{\partial
\tilde{\rho}}-\frac{2}{3} \tilde{\mu} \frac{\partial}{\partial
\tilde{\phi}}\Big)P_X[\tilde{Y}]+\nonumber
\\&& \frac{Q L}{4}\Delta_2 P_X[\tilde{Y}],
\end{eqnarray}
where $\Delta_2$ is Laplace operator. One can see that for an axially symmetric
distribution, i.e., when $P_X[X]$ depends only on $|X|=\rho$, the correction
(\ref{deltaPout2}) has the form $\delta P_{\mathrm{out}}[\tilde{\rho}]=\frac{Q
L}{4}\Delta_2 P_X[\rho]$, which is in agreement with the general (nonperturbative)
result, obtained in Ref.~\cite{Terekhov:2016a}, see Eq.~(32) therein. From Eq.
(\ref{deltaPout2}) one can see  that the first nonzero correction $\delta
P_{\mathrm{out}}[\tilde{Y}]$ to $P_{\mathrm{out}}[Y]$  has the order ${\cal
O}\left(\gamma Q L^2 \right)+{\cal O}\left(Q L/\rho^2\right)$, since $|Y|\sim |X|$.
Note that the validity of our approximation (\ref{deltaPout}) and the possibility to
use Laplace's method are justified by that the power $P$ is from the intermediate
power region  $ Q L \ll P \ll (\gamma^2 L^3 Q)^{-1}$: see the detailed explanation
in \cite{Terekhov:2016a}, Appendix C.

%=======================================
\section{Calculation of entropies}
To calculate the  conditional entropy with the accuracy $1/\mathrm{SNR}$ we
substitute  the conditional PDF (\ref{CondProbWithCorr})  to  Eq.
(\ref{condentropy}) and obtain:
\begin{eqnarray}\label{ExprForCondEntr1}
H[Y|X]&\approx &-\int{\cal D}X{\cal D}Y P_X[X]\Bigg(\log P_0[Y|X]\times\nonumber\\&& (P_0[Y|X]+\delta P_1[Y|X]+\delta P_2[Y|X])+\nonumber\\
&&\frac{\delta P_1^2[Y|X]}{2P_0[Y|X]}\Bigg).
\end{eqnarray}
To obtain Eq. (\ref{ExprForCondEntr1}) we used the consequence
(\ref{normalization3}) of the normalization condition  for the function $P[Y|X]$.
The direct integration over $Y$ in Eq.~(\ref{ExprForCondEntr1}) gives
\begin{eqnarray}\label{H[Y|X]}
H[Y|X]&\approx & H_0[Y|X]+\delta H[Y|X],
\end{eqnarray}
where
\begin{eqnarray}\label{H0[Y|X]}
H_0[Y|X] &=&1+\log(\pi Q L )+\nonumber\\ && \frac{1}{2} \int{\cal D}X P_X[X]\log\left({1+\frac{\mu^2}{3}}\right),\\
\delta H[Y|X]&=Q L&\int{\cal D}X P_X[X]\times\nonumber\\
&&\frac{\mu ^2 \left(-13\mu ^4+255 \mu^2+450 \right)}{150 \left(3+ \mu^2\right)^3
|X| ^2}.\label{dH[Y|X]}
\end{eqnarray}
The  leading in $1/\mathrm{SNR}$ term (\ref{H0[Y|X]}) for the conditional entropy
was obtained in Ref.~\cite{Terekhov:2016a}. Here we obtain the correction
(\ref{dH[Y|X]}). One can see that the correction (\ref{dH[Y|X]}) is proportional to
$Q$ and $\gamma^2$. Therefore, it  vanishes for the linear case $\gamma= 0$.

To calculate the output signal entropy (\ref{entropies}) we substitute
$P_{\mathrm{out}}[Y]$,  Eq. (\ref{PoutWithCorrections}), to Eq. (\ref{entropies})
and obtain
\begin{eqnarray}\label{HY}
H[Y]&=&-\int{\cal D}Y\Big( P_X[\tilde{Y}]\log P_X[\tilde{Y}]+\nonumber\\
&&\delta P_{\mathrm{out}}[\tilde{Y}]\Big\{1+\log P_X[\tilde{Y}]\Big\}\Big).
\end{eqnarray}
Let us note that ${\cal D} Y=dy_1 dy_2={\cal D} \tilde{Y}=d\tilde{y}_1
d\tilde{y}_2$. The first term in the right-hand side of Eq. (\ref{HY}) coincides
with the leading order contribution obtained in Ref. \cite{Terekhov:2016a}, see
Eq.~(39) therein. That is nothing else but the input signal entropy $H[X]$. The
second term in the right-hand side of Eq. (\ref{HY}) is proportional to parameter $Q
L$. We can omit the unity in the  curly brackets in Eq. (\ref{HY}) owing to the
normalization condition for $P_{\mathrm{out}}[Y]=P_X[\tilde{Y}]+\delta
P_{\mathrm{out}}[\tilde{Y}]$: $\int {\cal D} Y P_{\mathrm{out}}[Y] =1$, and
therefore, $\int {\cal D} \tilde{Y} \delta P_{\mathrm{out}}[\tilde{Y}] = 0$.

%======================================================================
\section{Optimal input signal distribution}
To calculate the channel capacity (\ref{capacity1}) we should find the optimal input
signal distribution $P_{\mathrm{opt}}[X]$ which is defined as
\begin{eqnarray}
C=\max_{P_{X}[X]} I_{P_{X}[X]}=I_{P_{{\mathrm{opt}}}[X]}.
\end{eqnarray}
To find the optimal input signal distribution $P_{\mathrm{opt}}[X]$ normalized to
unity and with the fixed average power $P$ we solve the variational problem, see
Section III in Ref.~\cite{Terekhov:2016a}:
\begin{eqnarray}\label{variationeq}
\delta J[P_{X},\lambda_{1},\lambda_{2}]=0
\end{eqnarray}
with the functional $J[P_{X},\lambda_{1},\lambda_{2}]$ that reads
\begin{eqnarray}\label{functional}
 J[P_{X},\lambda_{1},\lambda_{2}]&=&H[Y]-H[Y|X]-\nonumber\\
&&\lambda_1\left( \int {\cal D}X
P_{X}[X] -1\right)- \nonumber \\
&& \lambda_2\left( \int {\cal D}X P_{X}[X] |X|^2 -P
\right),
\end{eqnarray}
where $\lambda_{1,2}$ are the Lagrangian coefficients, at that $H[Y|X]$  and $H[Y]$
are given by Eqs. (\ref{H[Y|X]}) and (\ref{HY}), respectively. The solution of the
equation (\ref{variationeq}) in the leading order in the parameter $Q$ was found  in
Ref. \cite{Terekhov:2016a}:
\begin{eqnarray}\label{optimalPDF}
&&P_{\mathrm{opt}}^{(0)}[X]=N_{0}\frac{\exp\left\{- \lambda_{0}
|X|^2\right\}}{\sqrt{1+\mu^2/3}},
\end{eqnarray}
where $\mu=\gamma L|X|^2$. The  functions $N_{0}=N_{0}(P)$ and
$\lambda_{0}=\lambda_{0}(P)$ are determined from the conditions:
\begin{eqnarray}\label{normalization1}
\int {\cal D} X P^{(0)}_{\mathrm{opt}}[X]= 2\pi N_{0}\int^{\infty}_{0} \frac{d\rho
\,  \rho \, e^{- \lambda_{0}\rho^2}}{\sqrt{1+\gamma^2 L^2 \rho^4/3}} =1,
\end{eqnarray}
\begin{eqnarray}\label{normalization2}
\!\!\int \!\! {\cal D} X P^{(0)}_{\mathrm{opt}}[X] |X|^2 \!\! = 2\pi N_{0} \!\!
\int^{\infty}_{0} \!\! \frac{d\rho \, \rho^3 \! e^{- \lambda_0
\rho^2}}{\sqrt{1+\gamma^2 L^2 \rho^4/3}} =P.
\end{eqnarray}
The solutions $\lambda_0$ and $N_0$  can be found numerically for any arbitrary
case. Note that the products $\lambda_0 P$  and  $N_0 P$  are the functions of
dimensionless nonlinearity parameter  $\tilde{\gamma}=\gamma P L/\sqrt{3}$ only. For
the  case of small nonlinearity parameter $\tilde{\gamma}$ the solutions have the
form:
\begin{eqnarray}\label{NandlambdaasSmallGamma}
\lambda_0(P)=\frac{1}{P}\left(1-2\tilde{\gamma}^2\right),\,\,
N_0(P)&=&\frac{1}{\pi P}\left(1-\tilde{\gamma}^2\right).
\end{eqnarray}
In the case of  sufficiently large parameter $\tilde{\gamma}$ such as $\log
\tilde{\gamma} \gg1$ using the results of Ref. \cite{Terekhov:2016a} one can obtain
the following asymptotics:
\begin{eqnarray}\label{Nandlambdaas}
 \lambda_{0}&\approx& \frac{1-\log \log( c \tilde{\gamma})/\log(c\tilde{\gamma})}{P
\log(c \tilde{\gamma})},
\\ N_{0} &\approx& \frac{\tilde{\gamma}}{\pi P}\log^{-1}\left[c \tilde{\gamma}/( \lambda_{0} P) \right],
\label{Nandlambdaas1}
\end{eqnarray}
where $c= 2  e^{-\gamma_{E}}$ and the accuracy of asymptotic estimates
~(\ref{Nandlambdaas}) and (\ref{Nandlambdaas1}) is ${\cal
O}(\log^{-2}(\tilde{\gamma}))$.

To calculate the corrections of the order of $Q$ to the solution (\ref{optimalPDF})
we substitute the optimal input PDF in the following form
\begin{eqnarray}\label{PoptExpansion}
P_{\mathrm{opt}}[X]\approx P_{\mathrm{opt}}^{(0)}[X]+P_{\mathrm{opt}}^{(1)}[X]
\end{eqnarray}
to  Eq. (\ref{functional}), where $P_{\mathrm{opt}}^{(0)}[X]$ is defined in Eq.
(\ref{optimalPDF}) and $P_{\mathrm{opt}}^{(1)}[X]$ is the first correction
proportional to $Q$. Then we keep terms which are proportional to $Q$ and obtain:
\begin{widetext}
\begin{eqnarray}\label{PoptCorrection}
P_{\mathrm{opt}}^{(1)}[X]=Q
L\Big(-\lambda_0^2|X|^2+\frac{2\lambda_0}{1+\mu^2/3}+\mu^2\frac{(-137 \mu^4+1095
\mu^2+4950)}{4050|X|^2(1+\mu^2/3)^3}\Big)P_{\mathrm{opt}}^{(0)}[X]-(\delta\lambda_1+\delta\lambda_2|X|^2)P_{\mathrm{opt}}^{(0)}[X],
\end{eqnarray}
\begin{eqnarray}\label{deltalambda1}
&&\!\!\!\!\!\!\!\!\! \delta\lambda_1=\frac{Q L/P}{750
\tilde{\gamma}^2(\tilde{\gamma}^2 (P \lambda_0-1)+P \lambda_0-\pi P N_0)}\Big\{ 16
P^4 \lambda^2_0 (\lambda_0-\pi  N_0)^2+\tilde{\gamma}^4\Big(P \lambda_0(-1370+1379 P
\lambda_0)-428 \pi P N_0 \Big)+\nonumber \\&& \tilde{\gamma}^2\Big( P^2
\lambda^2_0(685+16 P \lambda_0 [P \lambda_0-4])+\pi P^2   N_0 \lambda_0 (48 P
\lambda_0-257)-428 \pi^2 P^2 N^2_0\Big)\Big\},
\end{eqnarray}
\begin{eqnarray}\label{deltalambda2}
&&\!\!\!\!\!\!\!\!\!\delta\lambda_2=\frac{ \lambda_0 Q L/P}{750 (\tilde{\gamma}^2 (P
\lambda_0-1)+P \lambda_0-\pi P N_0)}\Big\{ \tilde{\gamma}^2\Big(685-347 P
\lambda_0(1+P \lambda_0)+428\pi P N_0\Big)+\nonumber \\&& P \lambda_0\Big(315 \pi P
N_0-P \lambda_0(299+16\pi P N_0)\Big) \Big\}.
\end{eqnarray}
\end{widetext}
Since $P_{\mathrm{opt}}^{(0)}[X]$ obeys the normalization conditions
(\ref{normalization1}) and (\ref{normalization2}), therefore, the correction
(\ref{PoptCorrection})  must obey  the following two conditions:
\begin{eqnarray}\label{PoptCorrectionCond1}
&&\int {\cal D}X P_{\mathrm{opt}}^{(1)}[X]=0\,,\\
&&\int {\cal D}X |X|^2P_{\mathrm{opt}}^{(1)}[X]=0.\label{PoptCorrectionCond1}
\end{eqnarray}
One can check that for $\delta\lambda_{1,2}$ from Eqs. (\ref{deltalambda1}),
(\ref{deltalambda2}) these conditions are fulfilled.

%==================================================================================
\section{Capacity in the next-to-leading order}
To calculate the channel capacity up to the terms proportional to $Q$ we substitute
the optimal input signal distribution in the form (\ref{PoptExpansion}) to the
mutual information (\ref{MI}) and obtain
\begin{eqnarray}
C=C_0+\Delta C\,,
\end{eqnarray}
where the leading order contribution  $C_0$ reads, see Eq.~(51) in Ref.~\cite{Terekhov:2016a}:
\begin{eqnarray}\label{C0}
C_0=\log\left(\mathrm{SNR}\right)+ \lambda_0 P-\log (\pi  N_0 P)-1\,,
\end{eqnarray}
and the required next-to-leading correction has the form
\begin{eqnarray}
&\Delta C= \dfrac{1}{\mathrm{SNR}}\Bigg\{ \pi  N_0
P\left[\dfrac{214}{375}-\dfrac{8}{375}\left(\dfrac{\lambda_0
P}{\tilde{\gamma}}\right)^2 \right]+\nonumber \\&+ \lambda_0
P\left[\dfrac{137}{150}+\dfrac{8}{375}\left(\dfrac{ \lambda_0
P}{\tilde{\gamma}}\right)^2  \right]-\dfrac{347}{750} (\lambda_0 P)^2 \Bigg\}.
\label{capacityNLO}
\end{eqnarray}
The term $\Delta C$ is the first nonvanishing correction to the capacity. One can
check that for small  parameter $\gamma  L^2 Q \ll 1$ the correction
~(\ref{capacityNLO}) is always small. Indeed, the expression in the curly bracket in
Eq.~(\ref{capacityNLO}) divided by $\tilde{\gamma}$ is limited for all
$\tilde{\gamma}$. This correction can be calculated numerically for arbitrary
parameter $\tilde{\gamma}$, and analytically for small and large $\tilde{\gamma}$.

First, let us consider the correction at small nonlinearity.  We  substitute the
parameters $\lambda_0$ and $N_0$ in the form  (\ref{NandlambdaasSmallGamma}) and
obtain:
\begin{eqnarray}\label{CapacityCorrAtSmallGamma}
\Delta C\,\approx
\frac{1}{\mathrm{SNR}}-\frac{1}{\mathrm{SNR}}\frac{\tilde{\gamma}^2}{3}.
\end{eqnarray}
Using this result and expansion  of the $C_0$ at small nonlinearity, see Eq.~(53) in
Ref.~\cite{Terekhov:2016a}, we can write  the capacity within our accuracy in the
form:
\begin{eqnarray}
C\approx \log(1+\mathrm{SNR})-\tilde{\gamma}^2-\frac{1}{\mathrm{SNR}}\frac{\tilde{\gamma}^2}{3}.
\end{eqnarray}
One can see that the nonlinear correction  is negative for small $\tilde{\gamma}$
and it reduces the result for the linear channel.

More interesting is to consider the correction to the  capacity at large power $P$.
For the case $\log(\gamma L P)\gg 1$ and $P\ll(\gamma^2 Q L^3)^{-1}$ we have the
simple representation:
\begin{eqnarray}\label{DeltaCLarge}
&&\!\!\!\!\!\Delta C\,\approx \frac{1}{\mathrm{SNR}} \frac{214}{375}\pi N_0 P .
\end{eqnarray}
Using the asymptotic formulae (\ref{Nandlambdaas1}), (\ref{Nandlambdaas}) for
quantity $N_0 $ we arrive at the expression
\begin{eqnarray}\label{DeltaCLarge}
\Delta C\,&\approx& \frac{\gamma L^2 Q}{\sqrt{3}}\times \nonumber
\\&&
\frac{214}{375}\left(\log\left( c \tilde{\gamma}\log( c
\tilde{\gamma})\right)+\frac{\log \log(c \tilde{\gamma})}{\log( c
\tilde{\gamma})}\right)^{-1}.
\end{eqnarray}
We take notice that this correction is suppressed as $\gamma L^2 Q$  instead of
$1/\mathrm{SNR}=QL/P$ and it decreases as $1/\log\tilde{\gamma}$ at large
$\tilde{\gamma}$. For large $\tilde{\gamma}$ the correction (\ref{DeltaCLarge}) is
positive, therefore, it enhances the capacity.

\begin{figure}[t]
\begin{center}
\includegraphics[width=7cm]{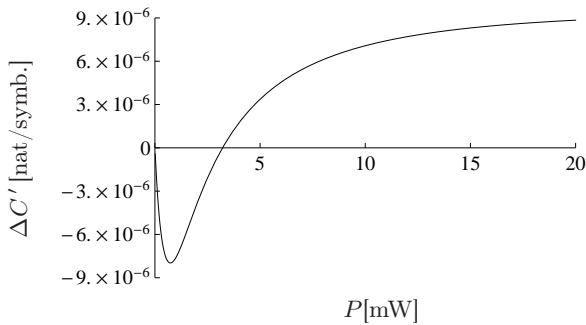} \hspace*{.4cm} \\
\begin{picture}(0,0)
\put(0,0){\text{$P$[mW]}} \put(-130,30){ \rotatebox{90}{\text{$\Delta C
\,'$\,[nat/symb.]}}}
%\put(0,120){\text{(a)}}
\end{picture}
\end{center}
\caption{\label{figure1}  The correction $\Delta C\,'$, see
Eq.~(\ref{CapacityNLOprime}), as a function of power $P$ for the parameters $Q=1.5
\times 10^{-7} \, \mathrm{mW}\, \mathrm{km}^{-1} $, $\gamma = 1.3 \times 10^{-3} \,
\mathrm{mW}^{-1} \mathrm{km}^{-1}$,  $L=1000 \,\mathrm{km}$.}
\end{figure}

\begin{figure}[t]
\begin{center}
\includegraphics[width=7cm]{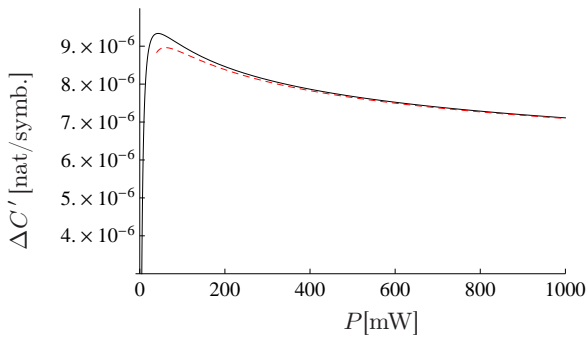} \hspace*{.4cm} \\
\begin{picture}(0,0)
\put(0,0){\text{$P$[mW]}} \put(-130,30){ \rotatebox{90}{\text{$\Delta C
\,'$\,[nat/symb.]}}}
%\put(0,120){\text{(a)}}
\end{picture}
\end{center}
\caption{\label{figure2} The correction $\Delta C\,'$, see
Eq.~(\ref{CapacityNLOprime}),  as a function of power $P$ for the parameters $Q=1.5
\times 10^{-7} \, \mathrm{mW}\, \mathrm{km}^{-1} $, $\gamma = 1.3 \times 10^{-3} \,
\mathrm{mW}^{-1} \mathrm{km}^{-1}$, $L=1000 \,\mathrm{km}$. The solid black line
corresponds to the exact expression obtained using Eq.~(\ref{capacityNLO}). The red
dashed line corresponds to the asymptotics obtained using Eq.~ (\ref{DeltaCLarge}).}
\end{figure}

For the further consideration of the correction it is convenient to subtract the
term $\frac{1}{\mathrm{SNR}}$, which corresponds to the expansion of the Shannon's
logarithm (\ref{CapacityShannon}) at large $\mathrm{SNR}$, from the correction
(\ref{capacityNLO}):
\begin{eqnarray}\label{CapacityNLOprime}
\Delta C\,'=\Delta C-\frac{1}{\mathrm{SNR}}.
\end{eqnarray}
The correction  $\Delta C\,'$ is convenient for analysis since it is regular
function for all range of signal power $P$. Let us consider the correction $\Delta
C\,'$ for the  parameters $Q=1.5 \times 10^{-7} \, \mathrm{mW}\, \mathrm{km}^{-1} $,
$\gamma = 1.3 \times 10^{-3} \, \mathrm{mW}^{-1} \mathrm{km}^{-1}$, $L=1000
\,\mathrm{km}$ which can be realized in experiment, see \cite{Mansoor:2011}. Note
that for chosen parameters the intermediate power region $QL\ll P\ll (\gamma^2 L^3
Q)^{-1}$ is extremely broad:
\begin{eqnarray}\label{regionnum}
1.5 \times 10^{-4} \mathrm{mW} \ll P \ll 3.89 \times 10^{3} \mathrm{mW}.
\end{eqnarray}
For these parameters the correction $\Delta C\,'$ and its asymptotics are  plotted
in Fig. \ref{figure1} and Fig. \ref{figure2} in the case of moderate  and  large
power $P$, respectively.
One can see that the correction $\Delta C\,'$  reaches the minimum $-7.97\times
10^{-6}\, $ nat/symb at $P \approx 0.73\, mW$ (it corresponds to
$\tilde{\gamma}_{min}\approx 0.55$), see Fig.~\ref{figure1}, and the maximum
$9.35\times 10^{-6}\,$ nat/symb at $P\approx 43.4\, mW$ (it corresponds to
$\tilde{\gamma}_{max}=32.82$), see Fig.~\ref{figure2}. In the wide power region $P
\gg (\gamma L)^{-1}\approx 0.76 \, mW$ and $P\ll (\gamma^2 L^3 Q)^{-1} \approx 4
\times 10^{3} \, mW$, see Fig.~\ref{figure2}, the correction $\Delta C\,'$ is almost
constant $\Delta C\,'\approx  4 \times 10^{-2} \times {\gamma Q L^2} \approx 8
\times 10^{-6}\,$ nat/symb.

%==========================================Conclusion
\section{Conclusion}

We  calculated the first nonzero  corrections to the optimal input signal
distribution $P_{\mathrm{opt}}$, the output signal distribution $P_{\mathrm{out}}$,
and channel capacity $C$ for the nondispersive nonlinear channel in the case when
the noise power $Q L$ is much less than the signal power $P$. These corrections are
proportional to the noise power $Q L$. We demonstrated that the correction $\Delta
C$ to the channel capacity is small in the intermediate power region $QL\ll P\ll
(\gamma^2 L^3 Q)^{-1}$. At large signal power $P$, $ (\gamma L)^{-1} \ll P\ll
(\gamma^2 L^3 Q)^{-1}$, the correction $\Delta C$ is the positive decreasing
function. We stress that $\Delta C$ is suppressed as $1/\mathrm{SNR}=QL/P$ for small
parameter $\tilde{\gamma}= \gamma L P/\sqrt{3}$ in comparison with the leading order
contribution, and it is suppressed as $\gamma L^2 Q$ decreasing as
$1/\log\tilde{\gamma}$ at large $\tilde{\gamma}$. The calculation of the channel
capacity $C_0$ was carried out in assumption that the parameter $\gamma^2 L^3 Q P
\ll 1$, or $P \ll (\gamma^2 L^3 Q)^{-1}$. Since among the corrections proportional
to $Q L$ there are no corrections of the order of $\gamma^2 L^3 Q P$ at large $P$,
we can expect that the next correction which contains power $P$ should be of the
order of $(\gamma^2 L^3 Q P)^2$, see Ref. \cite{Terekhov:2016a}. Therefore, the
applicability region at large $P$ for the channel capacity $C_0$ is determined by
the condition $(\gamma^2 L^3 Q P)^2 \ll 1$. For the given small parameter $\gamma^2
L^3 Q P$ this condition extends the applicability region for the channel capacity
$C_0$.

%==========================================Acknowledgments
\begin{acknowledgments}
The work was supported by the Russian Science Foundation (RSF) (grant No.
16-11-10133).
\end{acknowledgments}

%A.~V.~Reznichenko thanks the President program (СП-2415.2015.2) and the Russian
%Foundation for Basic Research (RFBR), Grant No. 16-31-60031/15 for support of the
%part of this work (asymptotics for the capacity at large power).

%===========================================Bibliography in PRE style

\end{document}